\def\a4{$a_4/a$}

\def\D4000{$\Delta$4000}

\def\etal{{\rm et al.}}

\def\lesssim{\,\lower2truept\hbox{${<\atop\hbox{\raise4truept\hbox{$\sim$}}}$}\,}
\def\gtrsim{\,\lower2truept\hbox{${>\atop\hbox{\raise4truept\hbox{$\sim$}}}$}\,}

\def\msole{~M_{\odot}}                  % simbolo di massa solare
\def\lsole{~L_{\odot}}                  % simbolo di lum solare
\def\msole{~M_{\odot}}                  % simbolo di massa solare
\input psfig.tex
\documentclass{aa}
\usepackage{graphics}
\begin{document}
\thesaurus{11.05.1; 11.05.2; 11.06.1; 11.06.2; 11.09.2; 11.19.3}
\title{Star Formation History of Early--Type Galaxies in Low Density 
Environments.}
\subtitle{V. Blue  line--strength indices for the nuclear region
\thanks{Table 2 is available in electronic form only, at CDS:
via anonymous ftp to
cdsarc.u-strasbg.fr (130.79.128.5) or via
http://cdsweb.u-strasbg.fr/Abstract.html}}

\author{M. Longhetti \inst{1}, A. Bressan \inst{2}, C. Chiosi
\inst{3}\thanks{Visiting Scientist, Max-Planck-Institut f\"ur Astrophysik,
K-Schwarzschild-Strasse 1, D-87540, Garching bei M\"unchen, Germany}, 
R. Rampazzo \inst{4}}

\offprints{M. Longhetti}

\institute{
$^1$Institut d'Astrophysique, 98 bis Boulevard Arago, 75014 Paris, France\\
$^2$Padua Astronomical Observatory, Vicolo dell'Osservatorio 5, I-35122 
Padua, Italy\\
$^3$Astronomy Department, University of Padua,  Vicolo dell'Osservatorio 5, 
I-35122 Padua, Italy \\
$^4$Brera Astronomical Observatory, Via Brera 28, I-20121 Milan, Italy
          }
\date{Received 1998; accepted }

\authorrunning{Longhetti et al.} 
\titlerunning{Star formation in early-type galaxies}

%%%%%%%%%%%%% Old Mode %%%%%%%%%
%  \markboth{}{Star formation in early-type galaxies}
%%%%%%%%%%%%%%%%%%%%%%%%%%%%%%%%%
   \maketitle

\begin{abstract}
We analyze the star formation properties of a sample of 21 shell galaxies and 30
early-type galaxies members of interacting pairs, located in low density
environments (Longhetti \etal\ 1998a, 1998b).

The study is based on new models developed to interpret the information coming
from `blue' H$\delta$/FeI, H+K(CaII) and \D4000 line-strength indices
proposed by Rose (1984; 1985) and Hamilton (1985).

We find that the last star forming event that occurred in the nuclear
region of shell galaxies
is statistically old (from 0.1 up to several Gyr) with respect to the corresponding
one in the sub-sample of pair galaxies ($<0.1$ Gyr or even ongoing
star formation).

If the stellar activity is somehow related to the formation of
shells, as 
predicted by several dynamical models of galaxy interaction,
shells have to be considered long lasting structures. 

Since pair members show evidence of very recent star formation,
we suggest that either large reservoirs of gas have to be present
to maintain active star formation,
if these galaxies are on periodic orbits, or
most of the pair members in the present sample are experiencing unbound encounters. 


\end{abstract}

\section{Introduction}
This paper is a sequel of a series dedicated to studying the star
formation history of early type galaxies showing shells, ripples, 
fine structures etc. which are taken as signatures of
past interactions (see Malin \& Carter
1983; Schweizer 1992; Reduzzi \etal\ 1996) and complex star
formation histories.

It is commonly accepted that dynamical interactions between galaxies
leading to the formation of shell structures are in general 
accompanied by star formation if enough gas is at disposal to the
interacting system. However, there are
different views of this subject that emerge from dynamical models.
In fact, numerical simulations of merging/accretion events 
with SPH codes, in which shell structures can develop, yield
contrasting predictions as far as the accompanying stellar activity is
concerned.
The models by Kojima \& Noguchi (1997) 
suggest that star formation in the satellite
is turned off well before the shells develop. This explains
why we see in shell galaxies signatures of post-star-burst.
In contrast, the models by
 Weil \& Hernquist
(1993) show that star formation occurs (in the galaxy center) while
the shell structure is forming.
In both cases the shell structure lasts for time scales shorter than
$\approx 1$ Gyr.

In alternative to the merging/accretion hypothesis (cf. Barnes 1996
and references therein), dynamical simulations show that shell
structures could also originate from 
 ``weak interaction''
between galaxies (Thomson 1991 and reference therein). In this case,
shells are expected to last much longer than with the previous
alternative.

The first paper of the series (Longhetti et al. 1998a) presented the 
sample of galaxies to be examined, i.e. 21 shell galaxies and 30 
members of isolated interacting pairs located in low
density environments and derived for their nuclear region several 
line strength indices 
defined in the spectral range  3700 \AA\ $< \lambda < $5700 \AA. The indices
were further splitted in two groups: {\it red indices} 
($\lambda >$ 4200 \AA) and {\it blue
indices} ($\lambda <$ 4200 \AA). The red indices (H$\beta$, Mg1, 
Mg2, etc..., 16 indices in total) 
are according to the definition by Worthey (1992) and Worthey et al.
(1994) and have been transformed into the Lick-IDS standard system, 
whereas the blue indices
are \D4000 \AA, H+K(CaII)
and H$\delta$/FeI according to the definition by Rose (1984, 1985),
Leonardi \& Rose (1996) and Hamilton (1985).
Finally, the spectra used to calculate the
indices had a 2.1 \AA\ FWHM resolution.

In the second paper (Longhetti et al. 1998b) the kinematics of the
gaseous and stellar components of the galaxies under investigation
were properly studied to correct the line-strength indices for
velocity dispersion.

With the aid of this material, a thorough comparison of the above indices
with those predicted by the fitting functions of several authors 
(Worthey 1992; Gorgas et al. 1993; Worthey et al. 1994; Idiart \& de
Freitas Pacheco
al. 1995; Buzzoni et al. 1992, 1994) was made by Longhetti et al. (1998c) in a
subsequent study. 
Furthermore, the properties of the
galaxies in the sample were compared with those obtained by Gonzalez
(1993), with particular attention to the distribution on the H$\beta$
vs. [MgFe] plane. The main result was that no difference could be
detected in this plane between interacting and post-interacting
galaxies. In particular, Longhetti \etal\ (1998c) discussed the influence of
a secondary burst of star formation induced by interaction/accretion
events.

In the present paper we will investigate the capability of the blue 
indices in unraveling the strength and age of the last star forming
episode. Indeed determining the strength and age of this episode 
would bear very much on our understanding of the 
formation of shell galaxies and the relationship between shell formation and 
star formation events.

The paper is organized as follows. Section 2 summarizes the basic properties of the
``blue'' indices. Section 3 presents models of Simple Stellar
Populations (SSP)
and describes how we calculate the line--strength indices by
amalgamating the medium resolution empirical library of 
stellar spectra by Jacoby \etal\ (1984, J84) with the library of
low resolution theoretical spectra
by Kurucz (1992). In addition to this, section 3 compares
the line strength indices directly calculated from spectra with those
obtained from fitting functions (see also Longhetti et al. 1998c).
Section 4 presents the line--strength indices for composite stellar
populations 
(CSP), i.e. model of galaxies in which a young stellar component is
added
to an old one in different proportions. This would mimic a recent burst
of stellar activity taking place in an old galaxy (Leonardi \& Rose
1996). The effects of emission lines and metallicity on the intensity
of the blue indices are also discussed in some detail.
The results of these simulations and how they compare with other
similar models in literature are discussed in section 5. Finally, some 
implications for dynamical models of shell formation and galaxy
evolution are discussed in section 6.

\section{The Line--Strength Indices}

\subsection{H+K(CaII)}
According to its definition (see Longhetti et al. 1998a), the index 
H+K(CaII) is actually a measure of the
Balmer H$\varepsilon$ line. It is in fact only the blend of H$\varepsilon$
with the H(CaII) line that is able to affect the apparent ratio
between H and K(CaII) lines. 

Rose (1985) has pointed out that stars of type later than F5 show a constant
ratio between H and K(CaII) lines, while hotter stars, with a deeper Balmer
series, are characterized by a stronger blend H(CaII)+H$\varepsilon$ relative to
K(CaII) and consequently by a smaller value of the H+K(CaII)index,
which reaches the minimum value in A-type stars 
characterized by the maximum Balmer absorption.

\subsection{H$\delta$/FeI}
This index, introduced by Rose (1985), is
a measure of the Balmer H$\delta$ line, whose central intensity
is related to the average intensity of two FeI lines.
At increasing surface temperature of a star, this index decreases
because the H$\delta$ and FeI lines increase and decrease,
respectively.
The H$\delta$/FeI index reaches its minimum value in A0-type stars
(whereby the Balmer absorption lines have their maximum intensity).
It turns out to be a good age indicator (just like all the Balmer lines)
even though it is also sensitive to metallicity via the dependence on
the
FeI lines. In addition to this, like the
other Balmer lines indices, it might be contaminated by emission lines.

\subsection{\D4000}

The spectral break at 4000\AA\ is produced by two factors: the 
concentration of Balmer lines toward the asymptotic limit of 3650\AA\
and the increase in stellar opacity at
shortward of 4000\AA\ caused by ionized metals. 
The \D4000\ index decreases with increasing surface 
temperature, it depends upon the stellar gravity only for stars
cooler than F5 (Hamilton 1985) and is also sensitive to the chemical
composition (Van den
Bergh 1963, van den Bergh \& Sackmann 1965, Hamilton 1985).
This index provides information
about the stellar population at the {\it turn off}, blending 
metallicity and age effects (Worthey 1992). 
Furthermore, the \D4000\ index is useful
for studies of distant galaxies, because thanks to the large
band-passes adopted for its definition it can be measured with small errors also
on spectra at low signal to noise.
Finally, just like all other ``blue''
indices, cosmological recession shifts this index toward redder
spectral ranges
whereby modern CCDs are
more sensitive.

\section{Line-Strength Indices of SSPs}

Most sudies of evolutionary population synthesis (EPS) are based on
low resolution stellar spectra, see for instance Bressan et
al. (1994, 1996) who made use of the Kurucz (1992) library.
This fact prevents the direct measure of line-strength indices 
on the integrated spectral energy distribution (ISED) of stellar 
aggregates of any complexity, e.g. SSPs and galaxy models.
The problem is solved by making use of the so-called {\it fitting
functions} that express the intensity of the indices as a function
of three basic stellar parameters, namely effective temperature,
gravity and metal content (see Worthey 1992, Worthey et al. 1994). 
Knowing these quantitities for each
single star in the stellar mix, the integrated line strength
indices
can be easily calculated both for SSPs and galaxy models in a fully
consistent manner with their associated ISEDs, see Bressan et
al. (1996)
for all details. It goes without saying that this is possible 
only if the fitting functions are known. Unfortunately, there are
indices, for instance the blue ones, for which fitting functions are
not
available. Therefore one is forced to make use of spectra with the
required degree of resolution. This is the subject of the coming
section.

\subsection{Using empirical Medium-Resolution Spectra}

In general, the spectral energy distribution $sp_{\lambda}(t,Z)$,
 flux (in suitable units) as a function of the wavelength,
 of a SSP of any age $t$ and metallicity $Z$ is defined by 

\begin{equation}
sp_{\lambda}(t,Z)=  \int_{M_{min}}^{M_{max}(t)}{\phi(M) f_{\lambda}(M,t,Z)dM}
\end{equation}

\noindent 
where the product between the initial mass function (IMF) $\phi(M)$
and the monochromatic flux, $f_{\lambda}(M,t,Z)$ of a star of
initial mass $M$, metallicity $Z$ and age
$t$, is integrated along the isochrone
associated with the SSP in question.
The integration is performed from $M_{min}$, the minimum
mass of objects that can become stars, up to $M_{max}(t)$, which is the
maximum mass of the stars in the most evolved phase still contributing to the
total luminosity (see for details Bressan \etal\ 1994, 1996).

The function $f_{\lambda}(M,t,Z)$ indicates the adopted spectrum of
the
constituent stars of mass $M$, age $t$ and metallicity $Z$,
i.e. the adopted
spectral library as function of effective temperature, gravity and
metal
content (a trivial re-normalization of the spectra is required to
scale the flux emitted by a star according to its total bolometric
luminosity). This
library
has of course its own resolution which determines the final result.

For the purposes of this study, we have adopted the empirical library
of stellar spectra by J84, which contains fluxes in units of
$[erg~ cm^{-2}~ s^{-1} $\AA$^{-1}]$ with a mean resolution 
of 4.5 \AA\ (FWHM) for stars of different spectral type and luminosity class.
The spectra cover the range of wavelength 3500\AA\ -- 7500\AA\
and refer to stars with solar composition. 

To incorporate the J84 library in our population
synthesis algorithm, we proceed as follows:

(i) We assign to each star of the J84 catalog the
gravity and effective temperature according to the relationships by
Garcia-Vargas (1991).

(ii) As the J84 spectra cover only a limited spectral
range, we have extended them both at the short and long wavelength
side by means of the Kurucz (1992) spectra of the same composition,
gravity and effective temperature. The Kurucz spectra have been 
cut at 3510\AA\ and 7427\AA\ and after suitable re-scaling, patched
with the J84 spectra. 

(iii) The spectral ranges used to calculate the scaling factors are
[3510\AA--3550\AA] and [7390\AA--7427.2\AA]
at the blue and red side of the spectrum, respectively.
From a theoretical point of view, the blue and red scaling factors
should be the same. 
Actually, we have detected differences between the blue and red 
 scaling factors of about 26\%. Various reasons concur to build up
this difference:
(a) inaccuracy of
the synthetic spectra; (b) errors in the flux calibration by
J84; (c) some mismatch in the correspondence between
Kurucz (1992) and J84 spectra via the correspondence
in gravity and effective temperature.

The monocromatic flux of individual stars in eq.(1)
$F_{\lambda}(M,T,Z)$
now reduces to $F_{\lambda}(M,T,Z_{\odot})$ because the J84 
library is only for solar composition. However, the dependence on the
metallicity (chemical
composition) is not completely wiped out, because it still remains via
the metal dependence of the isochrones along which relation (1) is 
integrated. Though reduced, the effect of different
metallicities (chemical compositions) on the final results can be
noticed.

In order to compare indices derived from direct measurements on the
spectra with those obtained from fitting functions, we must now
degrade the spectra of our SSP and/or model galaxies (see below)
to the resolution of 8.2 \AA\ (FWHM) typical of
the Lick-IDS spectra. To this purpose we convolve our spectra with
a Gaussian distribution with $\sigma$= 2.92\AA.

Using the degraded spectra we calculate the line-strength
indices by means of one of the relations below 

\begin{equation}
 I_a = (1 - {F_{I\lambda} \over F_{C\lambda}})  \delta\lambda 
\end{equation}

\begin{equation}
 I_m = -2.5 log({F_{I\lambda} \over F_{C\lambda}})
\end{equation}

\noindent
as appropriate to whether we are dealing with atomic or molecular indices
(see Worthey 1992, Worthey et al. 1994 and Longhetti et al. 1998a
 for the meaning of the various 
symbols).

In several cases, the results we have obtained do not perfectly agree
with the standard Lick
standard system, but a small off-set is present.
Determining this off-set is not possible because there is no sample of
stars or spectra in common between the Lick and the J84 
library.

Finally, we have calculated indices for three large grids of SSPs, i.e.
at varying metallicity (Z=0.004, Z=0.02, Z=0.05), each grid
containing
30 values of the age in the interval 0.05 to 19 Gyr.

%%%%%%%%%%%%%%Figure 1 ----------------------------------- 
\begin{figure}
\resizebox{9cm}{!}{\includegraphics{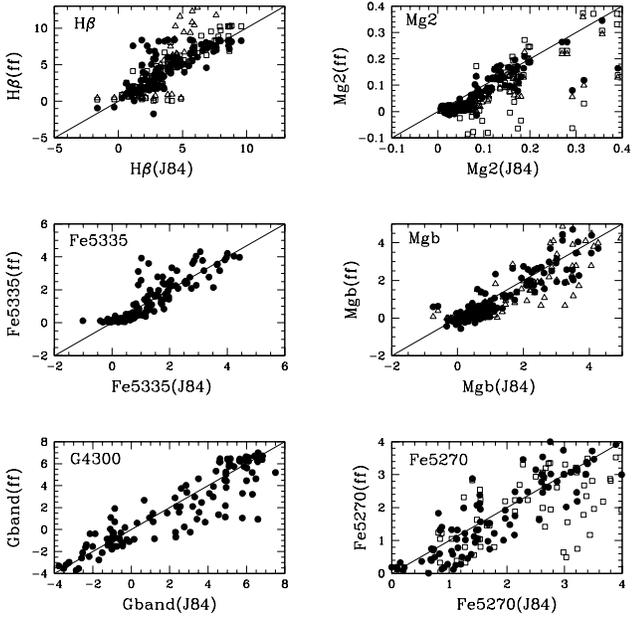}} 
\caption{Indices for individual stars: comparison between indices obtained from 
fitting functions (indicated with {\it ff}) and those derived
from direct measurements on
the stellar spectra of the Jacobi \etal\ (1984) library (indicated
with J84). The comparison with 
Buzzoni \etal\ (1992, 1994) (open squares) and Idiart \& de Freitas Pacheco (1995) 
(open 
triangles) is also shown}
\end{figure}
%----------------------end Figure 1 -------------------------------------

\subsection{Comparing stellar indices}
In this section we check whether stellar indices derived directly 
from the stellar spectra of the J84 library
(therein-after
the Jacobi indices) are fully
consistent with the same indices but obtained from fitting functions.
The comparison is made for all the stars of the J84 
catalog. 

In Fig.~1 we display the comparison between the Jacobi indices and
those 
obtained from the Lick system fitting functions (Worthey 1992, Worthey
et al. 1994) as, however, modified by Longhetti et al. (1998a,c) in the
high temperature regime.

Similar comparison has been made using the fitting function of
Buzzoni
 \etal\ (1992; 1994) for Mg2, Fe5270 and H$\beta$ (open
squares) and those of Idiart \& de Freitas Pacheco 
(1995) for Mg2, Mgb and H$\beta$. The results are also shown in
Fig.~1, where the
open squares are for Buzzoni et al. (1992, 1994) and open triangles for
Idiart \& de Freitas Pacheco (1995). Considering all the 
uncertainties affecting the
whole procedure, the agreement we get is remarkable.

%%%%%%%%%%%%%%Table 1 Index Comparison ***************
\begin{table} 
\caption[1]{Comparison of {\it red indices} derived from different
methods. 
The symbols {\it ff} and J84 stand for fitting functions and Jacobi
et al. (1994) library. See the text for more details. The SSP indices
are given at three ages (in Gyr) as indicated. }
\begin{center}
\begin{tabular}{l rr rr}
\noalign{\medskip} 
\hline
\multicolumn{5}{c}{Mean Value of [Index(ff) - Index(J84)]}\\
\noalign{\medskip} 
\hline
Index   & Stars &  SSP   & SSP    & SSP   \\
        & (J84) &($t>0.1$) &($t>1.0$) &($t>3.0$)\\
\noalign{\medskip}
\hline
H$\beta$   & 0.64  &  0.16  &  0.02  & -0.08 \\              
Mg1        &-0.01  &  0.02  &  0.03  &  0.03 \\
Mg2        &-0.02  & -0.01  &  0.04  &  0.05 \\
Mgb        & 0.24  &  0.26  &  0.51  &  0.58 \\
MgFe       &-0.24  &  0.14  &  0.31  &  0.36 \\
CN         &-0.07  & -0.01  & -0.01  & -0.01 \\
CN2        &-0.05  &  0.01  &  0.00  &  0.00 \\
Ca4427     & 0.07  &  0.10  &  0.17  &  0.20 \\
G4300      &-0.31  & -0.24  & -0.07  &  0.01 \\
Fe4383     &-0.28  &  0.21  &  0.53  &  0.63 \\
Ca4455     &-0.40  &  0.69  &  0.94  &  1.01 \\
Fe4531     &-0.22  &  0.03  &  0.16  &  0.23 \\
Fe4668     & 1.78  &  0.13  &  0.21  &  0.22 \\
Fe5015     & 0.70  &  0.43  &  0.57  &  0.59 \\
Fe5270     &-0.05  & -0.03  &  0.08  &  0.10 \\
Fe5335     & 0.63  &  0.22  &  0.25  &  0.27 \\
Fe5406     &-0.07  & -0.05  & -0.01  &  0.01 \\
\noalign{\medskip} 
\hline
\end{tabular}
\end{center}
%\label{tab_comp}
\end{table}

\subsection{Comparing SSP indices}

In this section firstly we examine the red indices for SSPs and compare
them
with their counterparts based on fitting functions. Secondly we look
at the blue
indices for SSPs. In such a case, no comparison with fitting function analogs is
obviously possible.

In Fig.~2 we show the comparison of SSP red indices with solar
metallicity ($Z$=0.02) obtained from the two methods (fitting
functions and direct measurement).
 Along each curve the symbols
change according to the age of the SSP: full squares for 10$\leq t
\leq$19 Gyr, open circles for
1$\leq t \leq$9 Gyr, full circles for 0.2$\leq t \leq$0.9 Gyr and
open squares for 0.05$\leq t \leq$ 0.1.

The data displayed in Fig.~2 show that both methods yield consistent
results. 
Although almost all red indices strongly depend on
the
metallicity, i.e. their values significantly change for small
variations in the metal content, this effect is not shown here as we
are
more interested now in the mutual consistency of the two methods
rather
than the absolute values of the indices as function of age and metallicity.

Particularly remarkable is the agreement for those indices that are
commonly considered as valuable age indicators, such as H$\beta$
and G4300.

Specifically, the values of H$\beta$ derived from the two methods
at ages older than
1 Gyr show an average difference of
$\approx 1\%$ with a root mean difference of $\approx 12\%$. 
For G4300 we find that they differ
by $\approx 8\%$ 
with a root mean difference $\approx 16\%$ passing from one method to
the other. Also in this case, the average
difference gets smaller for ages older than 1 Gyr (mean
difference $\approx 2\%$, root mean difference $\approx 6\%$).

%%%%%%%%%%%%%%Figure 2 ----------------------------------- 
\begin{figure}
\resizebox{9cm}{!}{\includegraphics{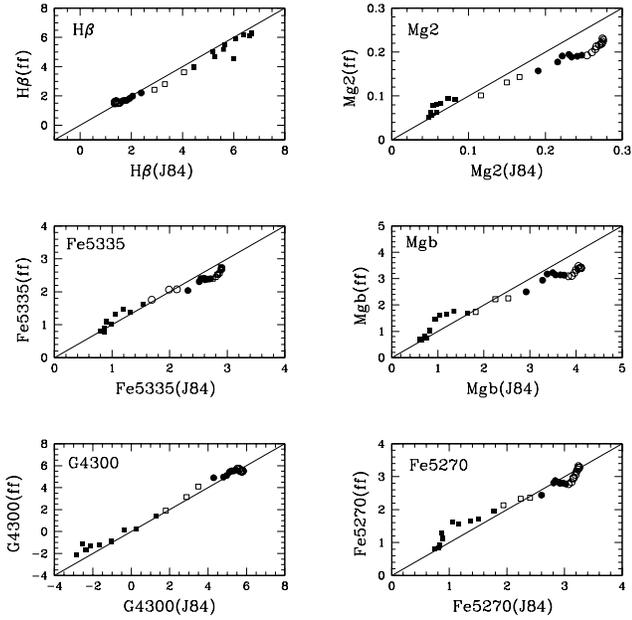}}
\caption{Indices for SSPs: comparison between indices in the Lick system
derived from fitting functions ({\it ff}) with those directly
measured on the spectrum
based on the Jacobi et al. (1984) library (J84).
The ages of SSPs are schematically indicated by 
full squares (10$\leq t \leq$19 Gyr), 
open circles (1$\leq t \leq$9 Gyr), full circles (0.2$\leq t \leq$0.9
Gyr) and open squares (0.05$\leq t \leq$ 0.1 Gyr). The metallicity of
the SSPs is solar ($Z=0.02$)}
\end{figure}
%----------------------end Figure 2 -------------------------------------

A systematic comparison between the Jacobi and Lick indices is
presented in Table~1 for all indices in common.

Finally, in the three panels of Fig.~3 we plot the blue indices 
\D4000, H+K(CaII) and H$\delta$/FeI as a function of the
age for SSP with different metal content. The effect of this 
is shown by using different symbols. The same data are also listed  in
Table~2 (available in electronic form only)
for the sake of general use. The following
remarks can be made:
 
(a) The indices H+K(CaII) and H$\delta$/FeI at ages older
than $2\div 3$ Gyr tend to flatten out or, at ages older than about
10 Gyr, even to reverse the trend. They somehow loose ability in 
deriving the age. The same is true for ages younger than 
0.3 Gyr. Therefore, there is only a limited range in age in which
these indices vary monotonically with time, i.e. 0.3 to about 3 Gyr.
However the effect of metallicity is not neglible bringing an uncertainty
on the age of about $\Delta log t \simeq 0.3-0.4$

(b) The index \D4000\ works much better than the previous ones,
because it is monotonic with the age up to ages older than 10 Gyr. The
uncertainty brought about by the metal parameter is however still
large ($\Delta log t \simeq 0.3-0.4$).

%%%%%%%%%%%%%%Figure 3 -----------------------------------
\begin{figure}
\resizebox{9cm}{!}{\includegraphics{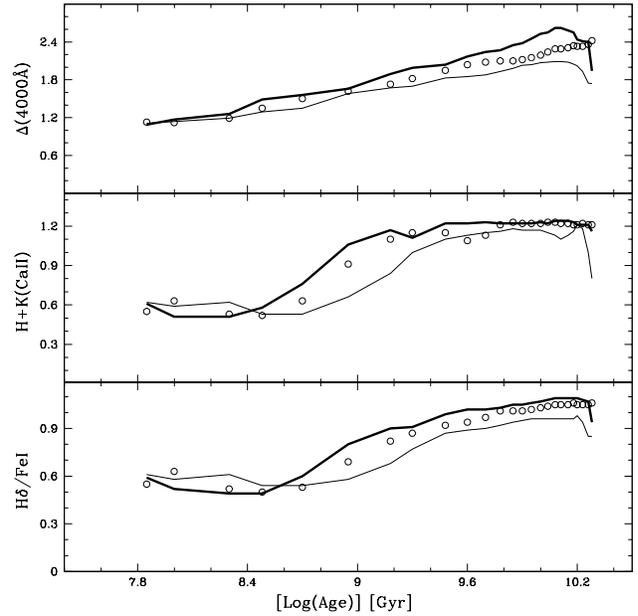}}
\caption{``Blue'' indices of SSPs as a function of the age ($t$).
The points refer to the solar metallicity ($Z=0.02$),
whereas the heavy solid line is for $Z=0.05$ and the thin solid line
is for $Z=0.004$. Limited to the case of solar composition, the age
intervals are put into evidence:
full squares (10$\leq t \leq$19 Gyr),
open circles (1$\leq t \leq$9 Gyr), full circles (0.2$\leq t \leq$0.9
Gyr),
full squares (0.05$\leq t \leq$ 0.1 Gyr)}
\end{figure}
%----------------------end Figure 3 ----------------------------------

%%%%%%%%%%%%%%%%%%%%%%Table 2 SSP Indices  ***************
% ONLY ELECTRONIC FORM 
\begin{table*}
\caption[2]{(available in electronic form only) The blue indices \D4000,  H+K(CaII)  and  H$\delta$/FeI as a function of the age for SSPs
 with different chemical composition.} 
\begin{center}
\begin{tabular*}{170mm}{l ccc cc ccc cc ccc}
\noalign{\medskip} 
 \hline 
\end{tabular*}
\end{center} 
\end{table*}
%%%%%%%%%%%%%%%%%%%%%%Table end end end  ***************

\section{Modeling Narrow Band Indices of CSPs}

If fine structures and/or signatures of interaction in early type
galaxies imply additional star forming episodes of various intensities
and ages, how they would reflect into the spectro-photometric properties 
of these systems? Do the line strength indices keep memory of this
past activity? How can we model this complex dynamical picture from
the point of view of spectro-photometry?

A viable approximation to the problem is to consider a 
galaxy of a certain mass, which underwent bulk star formation in the
far 
past, acquired its own pattern of abundances and ages for the mix of
stars, evolved passively ever since and at a certain age hosted an 
additional episode of star formation of suitable duration and
intensity in which a certain amount of gas was turned into stars.
The source of gas can be either internal, if any is left over and
retained
after the first initial activity, or acquired from outside during the
interaction episode.

If the initial star forming activity took place on a time
scale much shorter than the Hubble time, the host galaxy 
can be further approximated to a single entity, in
which stars span a certain range of age and metallicity.
In the
context of the standard galactic wind scenario (Larson 1974), the
bulk star forming activity is terminated within the first Gyr and the
mean metallicity of the stars (galaxy) is from nearly two times solar to a third
of 
solar as the galaxy mass decreases from 1 to 0.01 $\times 10^{12} M_{\odot}$
(see for instance the complete galaxy models
by Bressan et al. 1994, 1996; Tantalo et al. 1996, 1998).

The bursting mode idea has long been around starting from the 
pioneer studies by Huchra (1977) and Larson \& Tinsley (1978) to
explain the scatter in broad band colors observed in normal and
peculiar galaxies of the same
morphological type, till the more specific attempt by 
Leonardi \& Rose (1996) who investigated the effects of a burst of
star formation on the indices H+K(CaII) and H$\delta/FeI$.

Leonardi \& Rose (1996) start from the spectrum of a {\it normal}
galaxy, obtained as the average spectrum of about 70 elliptical galaxies
observed in the core of the Coma cluster and add to this template 
the synthetic spectra of young SSPs with solar metallicity and
different ages.

The age of the SSP stands for different burst ages, whereas the
burst strength is simulated by changing the relative proportions in
which
old and young spectra concur to build the final spectrum of the
{\it bursting galaxy}.

The indices H+K(CaII) and
H$\delta$/FeI are then measured on the composite spectrum 
of the post-star-burst models and
are compared with the observational data. The location of a sample of E+A
galaxies on the H+K(CaII) vs. H$\delta$/FeI diagram provides 
a first estimate of the ages and intensity of the star forming
activity characterizing E+A galaxies.

In this work we make use of the same technique.
The main difference with respect to Leonardi \& Rose (1996) is that
also for the {\it normal } galaxy we adopt a theoretical spectrum of suitable
age and metallicity. In all experiments we are going to present the
age of the quiescent
galaxy is 15 Gyr and the metallicity is solar.
Tests however have been made at varying the age and metallicity of the
 host galaxy.
Specifically, we have also adopted for the age the values of 12 and 18
Gyr and for the metallicity the values
$Z=0.004$ (20\% of the solar value) and $Z=0.05$ (2.5 times
the solar value). As expected the results are little affected by
 the choice for the age of the host galaxy, whereas they are more
sensitive to the metallicity.

%%%%%%%%%%%%%%%Figure 4 -----------------------------------
\begin{figure*}
\psfig{file=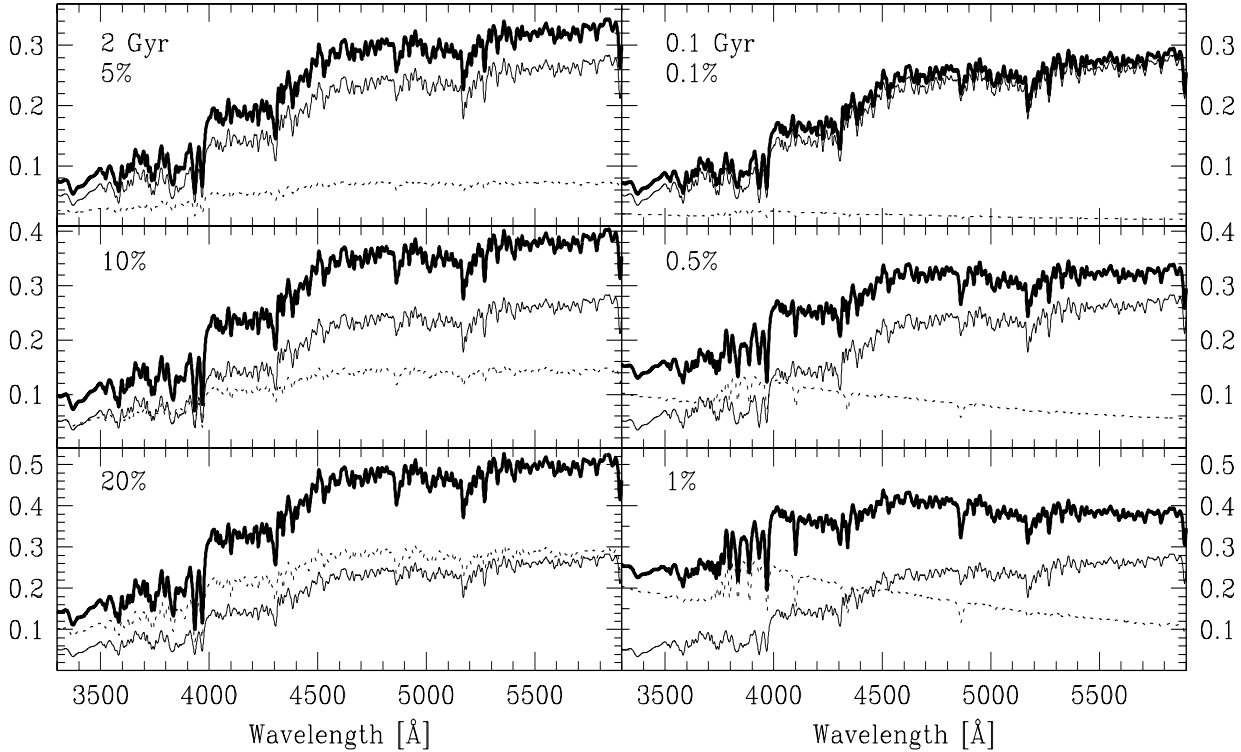,height=10.5truecm,width=16.9truecm}
\caption{Integrated spectral energy distribution (bold lines) of a 
mixed population made of an old (15 Gyr; thin solid line)
and a young SSP (dotted lines) component born in a recent
star forming event. Left panels refer to the case of a 
young component of 2 Gyr. 
The percentages of the mass involved in the 
star-burst episode are indicated at the top left of each sub-panel.
Right panels show the same but for a young component with age of 0.1 Gyr.
In all cases the metallicity is solar}
\end{figure*}
%----------------------end Figure 4 -------------------------------------

To simulate the bursts of star formation we have added to the template
spectrum of the host galaxy, spectra of SSPs of seven different ages,
i.e. 10, 6, 4, 2, 0.9, 0.1 and 0.05 Gyr.

In each simulation, the spectra of the two components (host galaxy and
SSP) are weighted by a factor expressing the percentage of stellar mass, 
relative to the total mass, created during the corresponding star
formation episode. 

Table~3 shows the correspondence between the fractionary mass in the
burst and the contribution to the total luminosity by the two
component. This
correspondence has been calculated using the mass to light ratio 
characterizing each SSP at the chosen age. 

The mass is given by the
total mass in living stars 
plus the mass in White Dwarfs and Neutron stars
for which we have assumed the typical values of 
0.6$\msole$ and 1.4$\msole$, respectively.
The dependence of the White Dwarf and Neutron Stars mass on the
progenitor mass is neglected here for the sake of simplicity.

The luminosity factor is
taken from the Padua library of SSPs (Bertelli et al. 1994, Bressan
et al. 1994, Tantalo et al. 1996).

%%%%%%%%%%%%%%Table 3 Mass - light ratios  ***************
\begin{table*} 
\caption[3]{Simulations of bursting
galaxies. $T$, $M_{star}$, $M_{bol}$ and $M/L$ are the age, mass and bolometric
magnitude and mass to light ratios 
of the SSPs representing the burst. $M_{star}$ and $M_{bol}$ refer to an 
ideal
SSP whose initial mass function in number is normalized to one over
the whole mass range.
The second half of the Table shows the ratio of the luminosity associated
to
the young SSP to the total luminosity of the system at varying the age
and mass percentage of the SSP representing the burst.}
\begin{center}
\begin{tabular}{lr    rrrrrrr}
\noalign{\smallskip}
\hline
\noalign{\smallskip} 
T [Gyr]             & 15.00 & 10.00 & 6.00 & 4.00 & 2.00 & 0.90 & 0.10 & 0.05 \\  
$M_{star} [\msole]$ &  0.95 &  1.06 & 1.21 & 1.36 & 1.69 & 2.38 & 5.81 &        \\  
$M_{bol}$           &  4.82 &  4.53 & 4.12 & 3.73 & 3.25 & 2.60 & 0.44 & -0.30 \\ 
$M/L$ $[\msole/\lsole]$ & 3.48 & 2.69 & 1.88 & 1.34 & 0.89 & 0.51 & 0.08 & 0.04  \\
\noalign{\smallskip}
\hline
\noalign{\smallskip}
\multicolumn{1}{l}{\% Y$_{SSP}$}&\multicolumn{7}{c}{$L_{Y_{SSP}}/L_{Total}$} \\
\noalign{\medskip}
\hline                           
0.01  &  &  0.01 &  0.02 &  0.03 &  0.04 &  0.08 &  0.56 &  1.10 \\ 
0.05  &  &  0.07 &  0.10 &  0.14 &  0.21 &  0.39 &  2.76 &  5.28 \\
0.1   &  &  0.13 &  0.19 &  0.27 &  0.42 &  0.77 &  5.38 & 10.03 \\ 
0.5   &  &  0.66 &  0.95 &  1.35 &  2.09 &  3.75 & 22.19 & 35.88 \\ 
1     &  &  1.31 &  1.89 &  2.68 &  4.12 &  7.27 & 36.44 & 52.94 \\ 
5     &  &  6.46 &  9.13 & 12.53 & 18.31 & 29.01 & 74.92 & 85.43 \\ 
10    &  & 12.72 & 17.50 & 23.22 & 32.12 & 46.31 & 86.31 & 92.52 \\
20    &  & 24.70 & 32.31 & 40.49 & 51.56 & 65.99 & 93.42 & 96.53 \\ 
30    &  & 35.99 & 45.00 & 53.84 & 64.60 & 76.89 & 96.05 & 97.95 \\ 
40    &  & 46.66 & 56.00 & 64.47 & 73.95 & 83.81 & 97.43 & 98.67 \\ 
50    &  & 56.75 & 65.62 & 73.13 & 80.98 & 88.59 & 98.27 & 99.11 \\
60    &  & 66.31 & 74.12 & 80.32 & 86.46 & 92.09 & 98.84 & 99.40 \\ 
70    &  & 75.38 & 81.67 & 86.39 & 90.86 & 94.77 & 99.25 & 99.62 \\ 
80    &  & 84.00 & 88.42 & 91.59 & 94.45 & 96.88 & 99.56 & 99.78 \\  
90    &  & 92.19 & 94.50 & 96.08 & 97.46 & 98.59 & 99.80 & 99.90 \\  
100   &  &100.00 &100.00 &100.00 &100.00 &100.00 &100.00 &100.00 \\  
\noalign{\smallskip}
\hline
\end{tabular}
\end{center}
%\label{tab_mass_light}
\end{table*}

The percentages of mass supposedly stored in the young SSPs ranges
from 0 (only the host galaxy is considered) to 100 (only the young
population contributes to the spectrum).

For purposes of illustration, we show in Fig. 4 a few cases
with different burst age and intensity (percentage of mass turned into
stars). Each panel contains the spectrum of the hosting galaxy, the
spectrum of the SSP representing the burst and the resulting total spectrum.
Worthy of note is that the contribution from the young component
grows at decreasing age of the burst and that even very small
percentages of
a young component may deeply alter the final spectrum. 
This is
particularly evident looking at the spectral region short-ward of 4000\AA, where, for
instance, the contribution by a SSP of 0.9 Gyr
involving only the 5\% of the total galaxy mass parallels that 
of the host galaxy, 15 Gyr old and containing the remaining 95\% of
the mass. 
Furthermore, a stellar
population of 0.1 Gyr representing only the 5\% of the total mass can produce
2/3 of the total luminosity emitted at $\lambda<4000$\AA. This comparison 
makes evident the usefulness of the spectral region short-ward of 4000\AA\
to unravel even small traces of recent star forming activity.

%%%%%%%%%%%%%%Table 4 Galaxy data ***************
\begin{table*} 
\caption[4]{Sample of shell and pair galaxies with best measured
H+K(CaII), \D4000 and H$\delta$/FeI}  
\begin{center}
\begin{tabular}{lccc c lccc}
\noalign{\smallskip}
\hline
\noalign{\medskip} 
\multicolumn{4}{c}{Shell Galaxies}& \multicolumn{4}{c}{Pair Galaxies}   \\
\noalign{\medskip}
\hline
\noalign{\medskip}
Name       &  \D4000\ & H+K(CaII) & H${\delta}$/FeI & & 
Name       &  \D4000\ & H+K(CaII) & H${\delta}$/FeI    \\
\noalign{\medskip}
\hline 
 E3420390  &             2.13    & 0.98      & 0.91             & &
 R101b	   &             2.15    & 1.10      & 0.87     \\  
 N1316	   &             2.12    & 1.15      & 0.89             & &
 R24b	   &             1.31    & 0.63      & 0.86     \\
 N6776	   &             2.04    & 0.82      & 0.94             & &
 R298b	   &             1.64    & 0.59      & 0.94     \\
 N7135	   &             2.19    & 1.09      & 1.03             & &
 R381a     &             2.28    & 1.23      & 0.92     \\
 E1070040  &             2.12    & 1.09      & 0.91             & &
 R405a	   &             2.26    & 1.02      & 0.96     \\
 E2890150  &             1.62    & 0.79      & 0.76             & &
 R405b     &             2.34    & 0.93      & 0.94     \\ 
 N1210	   &             1.83    & 0.47      & 0.96             & &
 R317a	   &             2.17    & 1.08      & 0.98     \\ 
 N6849	   &             2.27    & 1.00      & 0.94             & &
 R317b	   &             1.72    & 0.86      & 1.04     \\ 
 N6958	   &             2.02    & 1.16      & 0.93             & &
 R387a     &             2.16    & 1.60      & 1.01     \\
 N813	   &             2.20    & 1.11      & 0.85             & &
 R387b	   &             2.12    & 1.27      & 0.91     \\
 N1553	   &             2.27    & 1.22      & 0.97             & &
 R210a	   &             2.02    & 0.58      & 1.01     \\ 
 N1549	   &             2.26    & 1.22      & 1.00             & &
 R210b	   &             2.90    & 1.45      & 0.99     \\ 
 N1571	   &             2.16    & 1.15      & 0.94             & &
 R187b	   &             2.02    & 0.71      & 1.07     \\ 
 N2865	   &             1.69    & 1.12      & 0.80             & &
 R225a	   &             2.38    & 0.89      & 1.03     \\  
 N2945	   &             2.24    & 1.24      & 1.00             & &
 R225b	   &             2.50    & 0.80      & 1.00     \\
 N5018	   &             1.96    & 0.99      & 0.82             & &  
 R397b	   &             2.62    & 1.23      & 1.00     \\  
 E2400100a &             2.31    & 1.18      & 0.97             & &
           &                     &           &          \\
 E2400100b &             2.01    & 1.24      & 0.88             & &
           &                     &           &          \\
\hline
\end{tabular}
\end{center}
\label{tab_data}
\end{table*}

\section{The blue indices diagnostic}

In this section we compare the results of our simulations with the
data for real galaxies making use of two diagnostic planes. i.e. 
H+K(CaII) vs. H$\delta$/FeI and
H+K(CaII) vs. \D4000, displayed in Figs.5 and 6, respectively. 

The galaxy data are from the catalog of Longhetti et al. (1998a)
further selected by choosing only those objects with
 index H+K(CaII) measured
with less than 0.25 units of error. 
Due to the low S/N in the blue
part of the spectrum of some galaxies, only 18 shell galaxies and 16 pair
members pass the scrutiny. The basic data for this sub-sample of objects is
presented in Table~4.

The models (dashed lines) superposed to data in Figs.~5 and 6 refer to the solar
metallicity and show the expected
values for the three indices in post-star-burst galaxies. All lines
start from a common origin which is the locus in these diagrams of the
host quiescent galaxy (15 Gyr old and solar metallicity). The indices
of this reference galaxy are H+K(CaII)$\simeq$1.21,
H$\delta$/FeI$\simeq$1.12 and \D4000$\simeq$2.32. 
Each line refers to a
specific age of the burst event as annotated along the curves.
Moving
from the common origin, along each line what is varying is
the percentage of mass turned into stars during the burst event.
The percentage increases moving from the origin to the blue end of each
curve. This in fact represents the case of a 100\% young galaxy with
the age indicated.

In these diagrams, 
an old quiescent 
galaxy suffering from a burst of star formation would soon leave the
original place and move toward the blue corner
up to a maximum location dictated by the percentage of newly born
stars.
As soon as the burst is over and the new stellar component start
aging, the galaxy goes back to it original place. 
Once about 3 Gyr have elapsed, the current position is virtually 
indistinguishable from the original one (see also the run of the
three indices as a function of the age in Fig.~3).

The remarkable novelty showing up both in Figs.~5 and 6 is the
different
distribution of the shell and pair galaxies.
In fact, while shell galaxies fall in the region covered by 
the theoretical models, pair galaxies have a much broader
distribution.

The large error bars
affecting the data (in particular the index H$\delta$/FeI)
do not allow us to derive any specific conclusion about the age and
intensity of the secondary episode of star formation for individual
galaxies.
{\it The diagnostic diagrams have to be interpreted only in a statistical
sense}.

Recalling that old galaxies suffering from a
secondary burst of star formation perform the loop in the diagnostic
diagrams away from the quiescence position on a rather short time
scale (which in turn depends on the amount of mass engaged in the
burst), the different behavior of shell and pair galaxies 
can be explained if the bursting episode in the former is statistically older than
in the latter.

This is also somehow hinted by the large value of 
 H+K(CaII), larger than 1.2, that some pair galaxies reach. Such
 high
values of H+K(CaII) are expected only if the index is 
contaminated by the H$\varepsilon$ emission line (see below for more
 details). 
Therefore, the distribution
of pair galaxies in the two diagnostic planes is statistically indicative of
{\it ongoing star formation}. In contrast, the distribution of shell
galaxies seem to suggest that the secondary star forming episode is
on the average {\it older than 1 Gyr}.

We have checked that the above conclusions are not affected by the age
adopted for
the host galaxy at quiescence. Indeed similar results hold 
assuming for this object ages of 12 and 18 Gyr.

%%%%%%%%%%%%%%Figure 5----------------------------------- 
\begin{figure}
\resizebox{9cm}{!}{\includegraphics{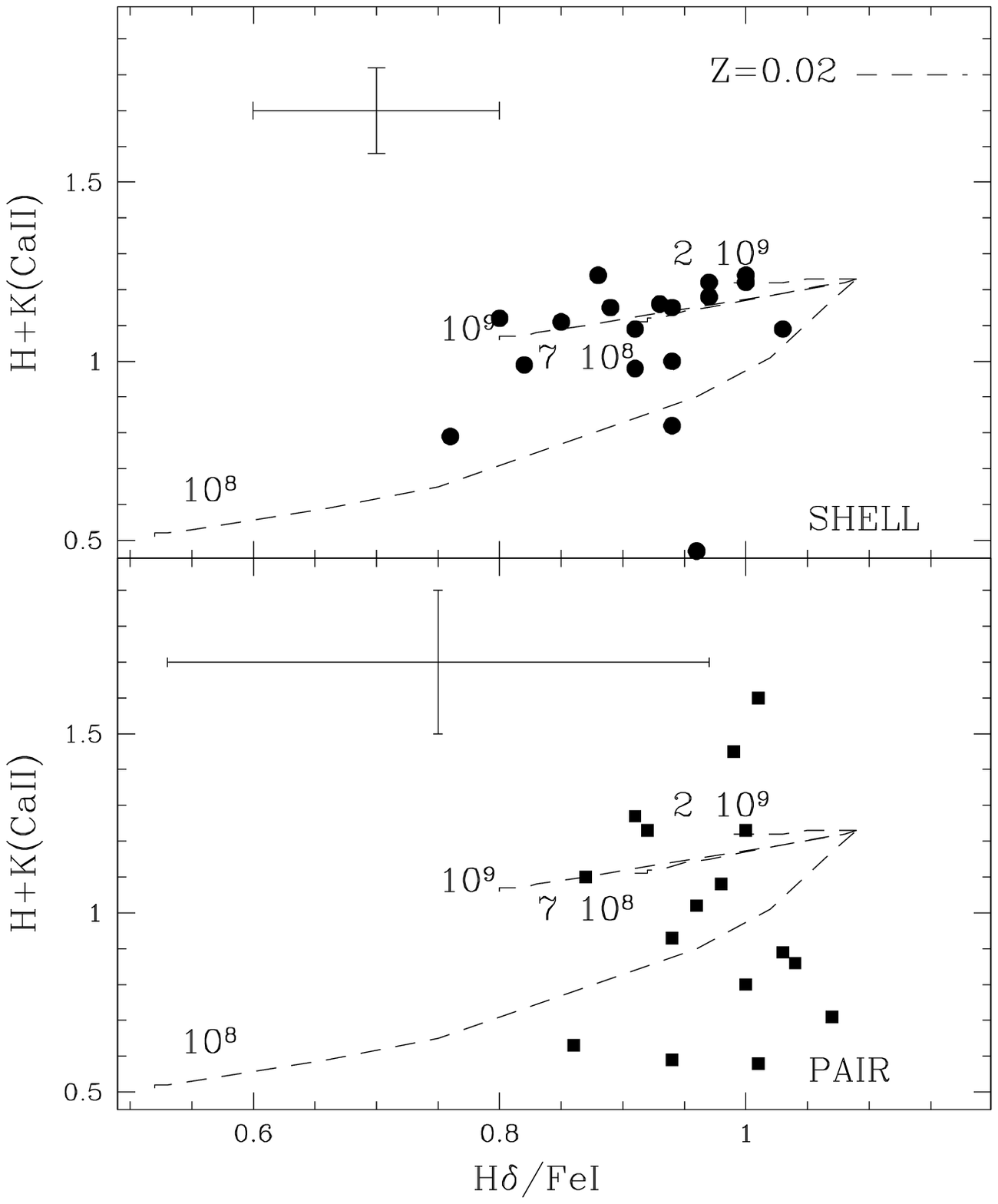}}
\caption{H+K(CaII) vs. H$\delta$/FeI diagram for shell galaxies (top panel) and
pair members (bottom panel). The dashed lines represent models of post-star-burst
galaxies with $Z=0.02$. Each line is labeled by the burst age. See
the text for
more details. In each panel the average error-bars are
shown} 
\end{figure}
%----------------------end Figure 5 ------------------------------------- 

%%%%%%%%%%%%%%Figure 6 -----------------------------------
\begin{figure}
\resizebox{9cm}{!}{\includegraphics{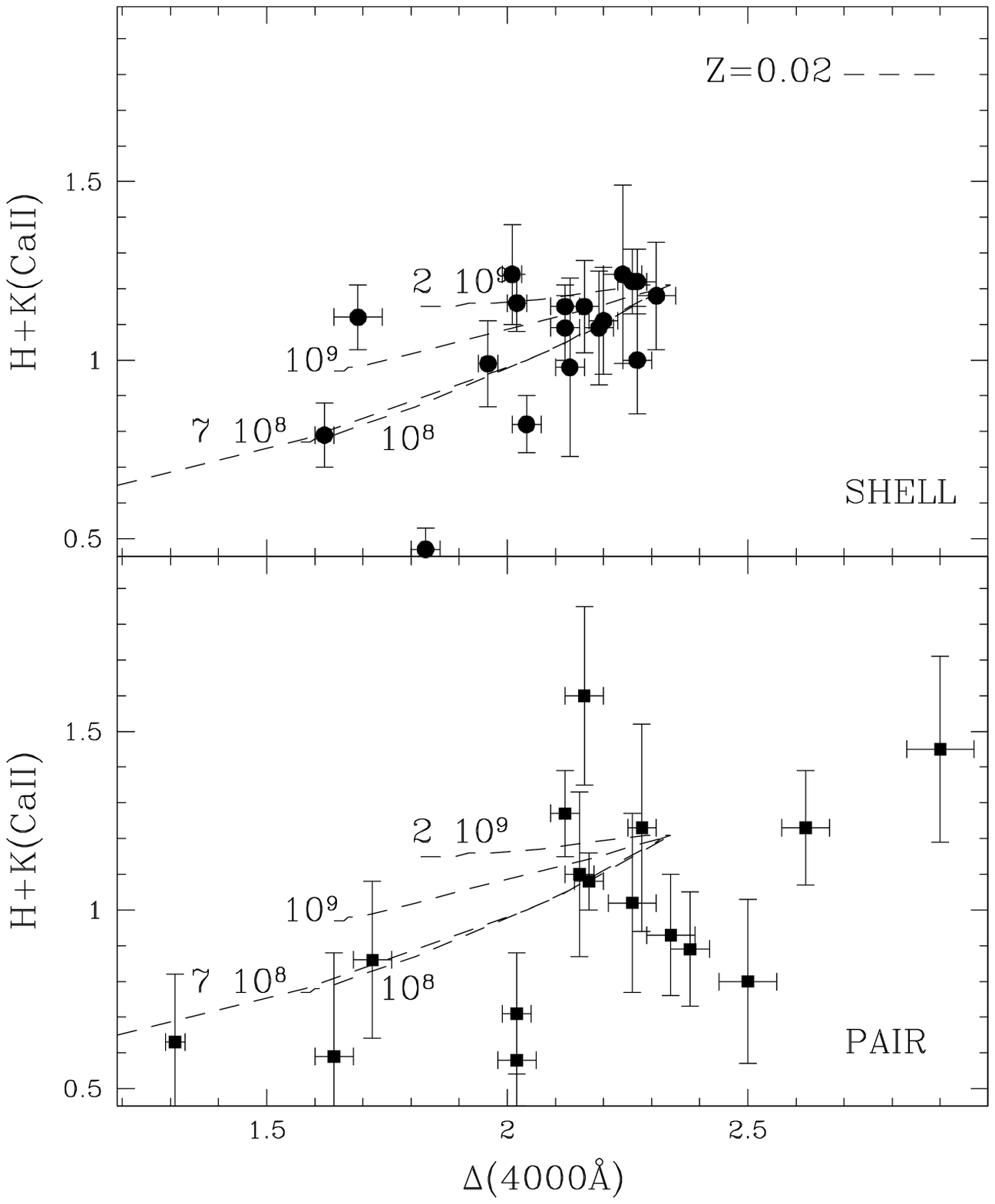}}
\caption{Same as Fig.5 but for the H+K(CaII) and \D4000\ indices}
\end{figure}
%----------------------end Figure 6 -------------------------------------

\subsection{Effect of Line Emission}

In order to support our preliminary suggestion that values of
H+K(CaII) higher than 1.2 are indicative of contamination by 
the H$\varepsilon$ emission line and hence ongoing star formation,
we have analyzed the spectroscopic atlas by Kennicutt (1992). 

It
contains
55 local galaxies, well distributed among the different
morphological types. Table~5 reports the results of our
measurements
 of the H+K(CaII) and \D4000\ indices for this sample. All peculiar and irregular
galaxies have been excluded from the analysis. 

The resolution of the
 Kennicutt (1992) spectra is between 5\AA\ and 8\AA\ FWHM, slightly better than
the one in usage here. While \D4000\ is not expected to
depend on the spectral resolution and/or velocity dispersion,
H+K(CaII) 
 tends to increase as the spectral lines get wider.
However, the expected increase in H+K(CaII), passing from the spectral
resolution of Kennicutt
(1992) data to ours, is very small and within statistical 
errors.

From the entries of Table~5, the mean value of the index
H+K(CaII) for
early--type galaxies (E+S0) is 1.19 ($\pm$ 0.16). In galaxies of
later types the index gets smaller. H+K(CaII) indeed 
decreases in stellar systems richer in A0-type stars, i.e. stellar systems
with active star formation. Interestingly enough, the value of
H+K(CaII) in post-star-burst models with solar (Figs.~5 and 6) 
and non solar metallicities
(not shown here for the sake of brevity) agrees with the maximum
observational
value $\simeq 1.2$.

In contrast, the bottom panels of Figs. 5 and 6 show some pair members
(RR210b, RR381a, RR387a, RR387b and RR397b) 
characterized by values of H+K(CaII)$\ge$ 1.3.
This means that the strength of the blend of H(CaII) with the Balmer
line H$\varepsilon$ is lower than expected. 

The only possible
explanation of this fact is to consider the contamination of the 
H$\varepsilon$
absorption line by the corresponding emission line. 

It is worth noticing that none of the above 5 galaxies shows a clear
H$\beta$ emission line.
Although the H$\varepsilon$ emission line is expected to have a lower absolute
intensity with respect to that of the H$\beta$ line, its influence on the H+K(CaII)
index could be greater than that of H$\beta$ on the corresponding absorption
feature. Then a small variation in the intensity of the H$\varepsilon$ absorption line 
caused by the corresponding emission component could be more easily
revealed in the H+K(CaII) index, thus making it a good indicator of
star formation.

\subsection{Effects of the Metallicity}

We have also calculated two grids of post-star-burst models with metallicities
different from the solar value. The results are shown in Fig.~7.
 As explained in \S~3, models of non solar
metallicity represent only a first order approximation because of the
incomplete
dependence of our simulations on this parameter (only in the isochrones and not in the
library
of stellar spectra). The new models are, however, much similar to the case of solar
metallicity presented above, even though they differ in some details.

(i) The reference galaxy of 15 Gyr age and the loci of the bursting
galaxies shift to bluer and redder indices passing from solar to
$Z=0.004$ and $Z=0.05$, respectively. In more detail, for $Z=0.004$ the
reference
galaxy has indices H+K(CaII)$\simeq$1.17, H$\delta$/FeI$\simeq$0.82
and \D4000$\simeq$2.05, whereas for $Z=0.05$ it has
H+K(CaII)$\simeq$1.23, H$\delta$/FeI$\simeq$1.18
and \D4000$\simeq$2.55. All the paths of the bursting galaxies are
accordingly shifted.
 
(ii) In the H+K(CaII) vs. H${\delta}$/FeI plane, 
different metallicities produce almost the
same loci for the post-star-burst galaxies.

(iii) A significantly larger variation is seen in the
H+K(CaII) vs. \D4000\ plane.

(iv) As compared to the observational data, the $Z=0.004$ models are too
blue, whereas the $Z=0.05$ ones are too red but still marginally
compatible.

Given the adopted approximation and
the uncertainty in the data (see the error-bars)
 we cannot contrive the mean metallicity of galaxies from their
location in the diagnostic planes 
 H+K(CaII) vs. H${\delta}$/FeI and H+K(CaII) vs. \D4000.
Nevertheless, the results shown in Figs.~5, 6 and 7 hint that the 
solar metallicity is perhaps 
suited to represent the mean value in early-type galaxies. 
Similar conclusions have been reached by Bressan et al. (1996),
Greggio (1996) and Longhetti et al. (1998c) analyzing the position of
early-type galaxies in the H$\beta$ vs. [MgFe] diagram.

%%%%%%%%%%%%%Table 5 ***************
\begin{table}
\caption{Data from the atlas of Kennicutt (1992)}
\halign{#\hfil&&\quad#\hfil\cr
\noalign{\hrule\medskip}
Class  & \D4000 & H+K(CaII) \cr
\noalign{\medskip}
\noalign{\hrule\medskip}
E+S0    &$2.08\pm0.23$&$1.19\pm0.16$ \cr
Sa      &$1.73\pm0.36$&$1.06\pm0.07$ \cr
Sb      &$1.57\pm0.11$&$0.91\pm0.10$ \cr
Sc      &$1.42\pm0.01$&$0.91\pm0.03$ \cr
\noalign{\medskip\hrule}}
\label{tab_kenn}
\end{table}
%********************  END    table 3 ****************

%%%%%%%%%%%%%%Figure 7 -----------------------------------
\begin{figure*}
\resizebox{16.9cm}{!}{\includegraphics{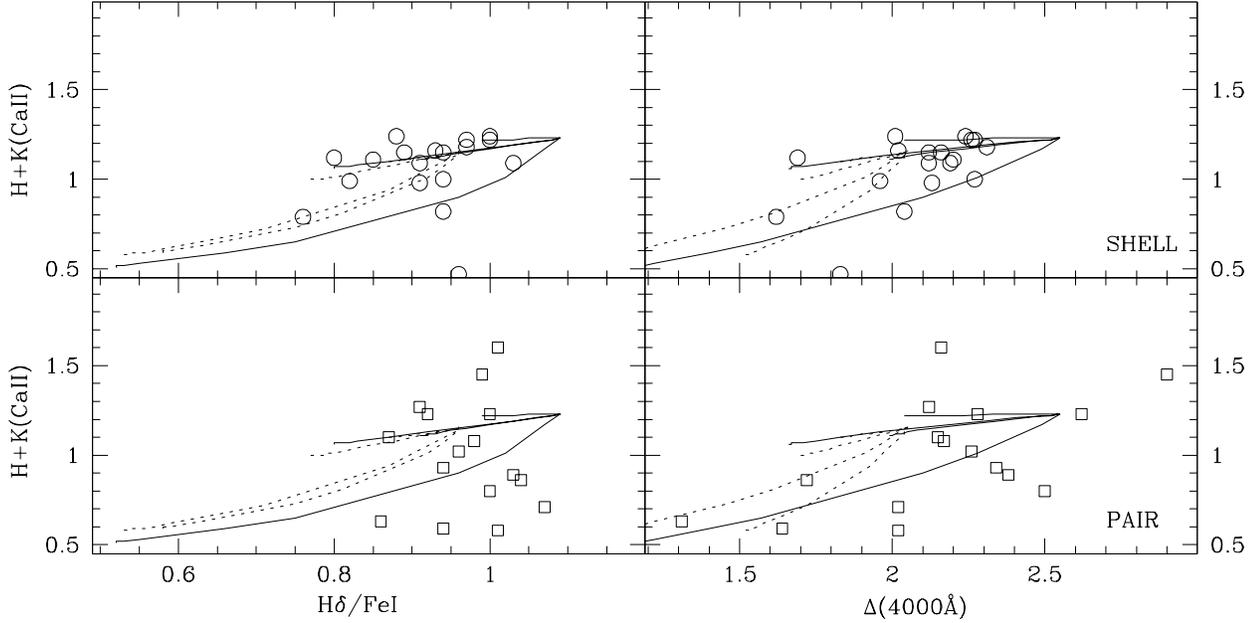}}
\caption{Effects of metallicity on the diagnostic diagrams H+K(CaII) vs.
 H$\delta$/FeI and H+K(CaII) vs. \D4000. The dotted and solid lines are for
 $Z=0.004$ and $Z=0.05$, respectively.}
\end{figure*}
%----------------------end Figure 7 -------------------------------------

\section{Summary and concluding remarks}

Adopting the medium resolution spectral library obtained by J84 
and the SSP models by Bressan \etal\
(1994, 1996) and Tantalo et al. (1996) we have
calculated three narrow band indices in the spectral range $\lambda
<$ 4200\AA, namely H+K(CaII), H$\delta$/FeI and \D4000, as
a function of age and metallicity.

With the aid of these results, we simulated post-star-burst galaxies by
means
of the simple recipe: an old host galaxy (with age of 15 Gyr and solar
composition) to which at some arbitrary age a burst of star formation 
is added. The intensity of this is measured by the percentage of mass
with respect to the total which is turned into stars by the burst
episode.

We have shown that the line strength indices H+K(CaII), H$\delta$/FeI and
\D4000\ respond to even small traces of past star formation activity. 
In particular, we have discussed the sensitivity of H+K(CaII) to 
the ongoing star formation events due to the contamination of the
Balmer H$\varepsilon$ emission line. 

From the analysis of the location of shell and pair galaxies on the
diagnostic planes based on the indices
H+K(CaII), 
H$\delta$/FeI and
\D4000\ we may advance some suggestions, whose validity is, however,
of mere
statistical nature owing to the large uncertainty still affecting the
data.

\begin{enumerate}

\item Shell galaxies and early-type galaxies members of pairs in the
diagrams 
show different
distributions in the \D4000\ vs. H+K(CaII) and H$\delta$/FeI vs. H+K(CaII) planes. 

\item In shell galaxies, the age of the last star forming event goes
from 0.1 to several Gyr and involves different percentages of mass.
If the last burst
of stellar activity that affects the line strength indices, 
correlates with the dynamical mechanism, i.e. merger or weak interaction,
forming the shell features, in such a case these latter are {\it long
 lasting phenomena}.

\item The distribution of early--type galaxies members of pairs
 suggests that the vast
majority of them contain a very young stellar component (between 0.1 and $\leq$1 Gyr), 
i.e. they suffer from a {\it very recent} burst of star formation.

\end{enumerate}

Let us now discuss the above results in the context of dynamical
models of galaxy interaction, having in mind that the formation of 
shell structures of long duration is a sort of constraint
hinted by the observational data.

Among the N-body and/or SPH dynamical models dealing with the formation of
shells, merger models are in general unable to produce shells of long duration. 
Dupraz \& Combes (1986) may produce shells that last
longer than in other merger models (Quinn 1984; Hernquist \&
Quinn 1989), depending on the initial conditions. However these models
fail in predicting the radial distribution of shells in
the best studied shell galaxies.
Only the 
inclusion of the dynamical friction (Dupraz \& Combes
1987), that slows down the merging process between the two galaxies, may
produce long lasting shells with the correct distribution. Nevertheless,
these models predict also
the presence of a double nucleus in the galaxy hosting the shells.
Noteworthy, in the
present sample of shell galaxies only ESO~240-100, out of twenty objects, is
characterized by the presence of a double nucleus.

Furthermore, SPH simulations of mergers by Weil \& Hernquist (1993) show that 
during the shell structure phase, a large fraction of the
gas content 
(if present) falls soon onto the nucleus whereby 
star formation is likely to occur. In these models
shells are as old as the last central star forming event. Similar predictions 
are found in the models by Kojima and Noguchi 
(1997), in which the time delay between the end of the star-burst
and the
shell formation is very short ($\leq1$ Gyr). This prediction
hardly matches the occurrence of shell structures and relatively old
star-bursts indicated by line strength diagnostic.

Alternatively, dynamical models producing shells via {\it weak
interaction} event (Thomson \&
Wright 1990, Thomson 1991) predict much longer lifetimes for the shells
up to $\approx 10$ Gyr. This type of model requires, however, 
the presence of a thick disc structure in the host early-type galaxy. 
Furthermore, shells do not develop if the
galaxies experience multiple interactions.

Rampazzo \etal\ (1998b) analyzed the isophotal structure of
pair galaxies in our sample and found that a large fraction of them show a 
disc-like 
structure. Therefore, if the interaction generating shells is 
the weak interaction mechanism of Thomson \& Wright (1990) and Thomson
(1991), we would expect to see them associated to pair galaxies where
the disc-like structure is often present. Among the 
pair galaxies of our sample, only a few possess shell structures,
namely RR~225a, RR~225b and RR~278a to which the system 
NGC~1549+NGC~1553 (the difference among the two systemic velocities is
$\Delta V$ 31 km~s$^{-1}$: Longhetti \etal\ 1998b) in the Dorado group
can be added. 

Finally, let us comment on the observational hint that despite the
large uncertainty affecting the indices we have derived, 
many pair galaxies seem to be characterized by very
young bursts.

Among others, RR278a is particularly remarkable as it has the
H+K(CaII) index as high as 1.95 and most likely H$\beta$ lines
in emission, suggesting ongoing star formation.
Another interesting case is NGC~1553 in the center of which 
Trinchieri \etal\ (1997), using narrow
band imaging, have found H$\alpha$ emission. Longhetti \etal\ (1998a,b)
did not detect H$\beta$ emission probably because it was too faint or it was filling
the absorption feature. If so, also the value of the H+K(CaII) 
line may be contaminated by H$\varepsilon$ in emission. It follows
from all this that burst in NGC~1553
could be as recent as in pair galaxies with shells. 

{\it What the implications are as far as the star formation history
 in pair galaxies is concerned?}

Prior to any other consideration, one has to clarify the question
whether the pair galaxies in question are bound or unbound systems.
The criteria used to select pairs of galaxies are designed to
isolate objects whose interaction potential energy is by far larger
than that with any other nearby galaxy. Nevertheless, there remains
the controversial problem of the relative percentage of 
bound vs. unbound pairs (see e.g.
Junqueira \& de Freitas Pacheco 1994 and reference therein). 

In recent years, efforts
have been made to disentangle the problem of the physical reality of
pairs by means of techniques different from the 
classical statistical methods. For instance, the X-ray diffuse
background has been proposed as a probe of the gravitational
potential well and 
physical reality of a pair/group in turn.

The NGC~2300 group, dominated by the pair 
NGC~2300/NGC~2276 (Mulchaey \etal\ 1993, 1996) and surroundend by 
X-ray diffuse background could be taken as an example of
a bound system. In contrast, the absence of an
X-ray diffuse background in K~416 pair of galaxies (Rampazzo \etal\
1998a), 
the analog of NGC2300 as far as morphology is concerned, suggests this
system is an unbound encounter.

In this context, the
fact that the majority of pair members in our sample underwent a recent
star formation event (in some of them even going on now) suggests that either 
large reservoirs of gas must be present to maintain active star formation
(if these galaxies are bound systems on periodic orbits) or most
of the pair galaxies in the sample are experiencing unbound 
encounters and undergo an occasional (perhaps unique) burst of stellar
activity.

\begin{acknowledgements}
ML acknowledges the kind hospitality of the Astronomical Observatories
of Brera (Milan) and Padua during her PhD thesis and
the support by the European Community under TMR grant ERBFMBI-CT97-2804.
RR acknowledges the hospitality of the Asiago and Padua Astronomical
Observatories where part of this study has been carried out. 
CC whishes to acknowledge the 
friendly hospitality and stimulating environment provided by MPA in
Garching
where this paper has been completed during leave of absence from the
Astronomy
department of the Padua University.
ML, AB and CC acknowledge the support by the European Community
under TMR grant ERBFMRX-CT96-0086.
\end{acknowledgements}

\end{document}